\documentclass[aps,prd,showpacs,preprintnumbers,superscriptaddress,nofootinbib]{article}
\usepackage[dvips]{graphicx}
\usepackage{bm,latexsym,amsmath,amssymb,amsfonts}
\usepackage{amsmath,amssymb,amsfonts,amsthm}
\newtheorem{prop}{Proposition}

\newcommand{\R}{\mathbb R}

\newcommand{\nablash}{\nabla{\kern -.75 em
     \raise 1.5 true pt\hbox{{\bf/}}}\kern +.1 em}
\newcommand{\Deltash}{\Delta{\kern -.69 em
     \raise .2 true pt\hbox{{\bf/}}}\kern +.1 em}
\newcommand{\Rslash}{R{\kern -.60 em
     \raise 1.5 true pt\hbox{{\bf/}}}\kern +.1 em}

\newcommand{\eps}{\varepsilon}

\newcommand{\RN}{Reissner-Nordstr\"om}

\begin{document}

\title{A counterexample to a Penrose inequality conjectured by Gibbons  }

\author{Sergio Dain$^{1,2}$,  Gilbert Weinstein$^3$ and  Sumio Yamada$^4$
 \\
  $^1$Facultad de Matem\'atica, Astronom\'{i}a y F\'{i}sica, FaMAF,\\
  Universidad Nacional de C\'ordoba,\\
  Instituto de F\'{\i}sica Enrique Gaviola, IFEG, CONICET,\\
  Ciudad Universitaria,  Argentina.  \\
  $^{2}$Max Planck Institute for Gravitational Physics,\\
  (Albert Einstein Institute), Germany. \\
  $^{3}$ School of Mathematics, Monash University\\
  $^{4}$ Mathematical Institute\\
  Tohoku University, Japan\\}


\maketitle

\begin{abstract}
  We show that the Brill-Lindquist initial data provides a counterexample to a
  Riemannian Penrose inequality with charge conjectured by G.~Gibbons. The
  observation illustrates a sub-additive characteristic of the area radii for
  the individual connected components of an outermost horizon as a lower bound
  of the ADM mass.

\end{abstract}

\section{Introduction}
Let $(M, g_0)$ be an asymptotically flat three dimensional manifold, with 
nonnegative scalar
curvature $R_0$.  Given an asymptotically flat end, let us assume that there
exists a set of minimal 2-spheres acting as the outermost horizon. In this
situation, there are a series of inequalities which relate the asymptotic data
with the Riemannian geometry of the manifolds.

The first such inequality is the Positive Mass
Theorem~\cite{Schoen79b,Schoen81,Witten81}.  We rephrase the Riemannian
version of this result as the following variational statement:
\emph{among all time-symmetric asymptotically flat initial data
sets for the Einstein-Vacuum Equations, flat Euclidean 3-space is
the unique minimizer of the total mass.}  Thus, the total mass
satisfies $m\geq0$ with equality if and only if the data set is
isometric to $\R^3$ with the flat metric.  

A stronger result is the Riemannian version of the Penrose Inequality
\cite{Penrose73, Bray, Huisken01}(see also the review article \cite{Mars:2009cj} and references
therein), which can be stated in a similar variational vein: \emph{among all
  time-symmetric asymptotically flat initial data sets for the Einstein-Vacuum
  Equations with an outermost minimal surface of area $A$, the Schwarzschild
  slice is the unique minimizer of the total mass.} In other words, $m\geq R/2$
where $R=\sqrt{A/4\pi}$ is the area radius of the outermost horizon, and
equality occurs if and only if the data is isometric to the \emph{Schwarzschild
  slice}:
\[
    g_{ij} = \left(1+\frac{m}{2r}\right)^4  \delta_{ij}.
\]

When these results are phrased in this fashion, a natural question
is whether similar variational characterizations of the other
known stationary solutions of the Einstein equations hold. In
particular, one could ask whether among all asymptotically flat
Einstein-Maxwell initial data set   with an horizon of
area $A$ and charge $Q$  the \RN\ slice is the unique minimizer of the mass. 
 This is equivalent to asking whether
the following inequality holds for any data set:
\begin{equation}    \label{eq:ChargedPenrose}
    m \geq \frac12 \left(R + \frac{Q^2}{R} \right),
\end{equation}
where $Q$ is the total charge, with equality if and only if the
data are a \RN\ slice. The charge within a $2$-surface $S$ is defined by 
\begin{equation} \label{eq:charge}
    Q(S) = \frac1{4\pi} \int_S E_i n^i \, dA
\end{equation}
It depends only on the homological type of $S$.

When the horizon is connected,
inequality~\eqref{eq:ChargedPenrose} can be proved by using the
Inverse Mean Curvature flow~\cite{Huisken01,Jang79}. Indeed,
the argument in~\cite{Jang79} relies simply on Geroch monotonicity of
the Hawking mass --- which still holds for the weak flow
introduced by Huisken and Ilmanen in~\cite{Huisken01}
--- while keeping track of the scalar curvature term
$R=2\bigl(|E|^2+|B|^2\bigr)$. However, when the horizon has
several components the same argument yields  the following
inequality:
\begin{equation}
  \label{eq:1}
    m \geq \frac12 \max_i \left( R_i + \frac{\left(\min \sum_i \eps_i
    Q_i\right)^2}{R_i} \right),
\end{equation}
where $R_i$ and $Q_i$ are the area radii and charges of the
components of the horizon $i=1,\dots,N$, $\eps_i=0$ or $1$, and
the minimum is taken over all possible combinations.

In \cite{Weinstein05} we pointed out
that \eqref{eq:ChargedPenrose} does not hold for the case of an horizon with several disconnected
components. Namely, there exists a strongly asymptotically flat time-symmetric initial
data set $(M,g,E,0)$ for the Einstein-Maxwell Equations such that:
\begin{equation}    \label{eq:ViolateChargedPenrose}
    m < \frac12\left(R + \frac{Q^2}{R}\right).
\end{equation}

In 1984, Gibbons~\cite{Gibbons84} conjectured an inequality similar 
to~\eqref{eq:ChargedPenrose}. However, in his conjecture, the
right hand side of~\eqref{eq:ChargedPenrose} is taken to be
additive over connected components of the horizon.  Thus,
Gibbons's conjecture states that:
\begin{equation}    
\label{eq:GibbonsPenrose}
    m \geq \frac12 \sum_i \left(R_i + \frac{Q_i^2}{R_i}\right).
\end{equation}
There is a physical reason to introduce additive quantities on the right hand side of the
inequality. The quantity
\begin{equation}
  \label{eq:2}
  m_i= \frac12 \left(R_i + \frac{Q_i^2}{R_i}\right),
\end{equation}
appears to play the role of the quasi-local mass of the $i$-th black
holes. Then, inequality \eqref{eq:GibbonsPenrose} can be interpreted as saying that the total mass of the spacetime
is always bigger than the sum of the quasi-local
masses of the individual black holes.   There is, however, a
Newtonian reasoning to doubt inequality \eqref{eq:GibbonsPenrose} in the case of
several black holes. When the black holes are separated by a large distance, it is
expected that the interaction energy between them, which is asymptotically Newtonian in this
limit, will be negative. The total energy of the spacetime should be the sum of
the quasi-local masses of the black holes plus this negative interaction
energy. Hence the sum of the quasi-local masses is expected to be bigger than the
total mass. The counterexample we present in the next section exhibits precisely
this behavior. 

It is important to note that there is an inequality analogous to 
\eqref{eq:GibbonsPenrose} for the Kerr black hole:
\begin{equation}
  \label{eq:3}
  m\geq \sum_i  m_i.
\end{equation}
where the quasi-local masses are now defined by:
\begin{equation}
  \label{eq:32}
  m_i=\sqrt{\frac{R^2_i}{4}+ \frac{J^2_i}{R^2_i}},
\end{equation}
Here, $J_i$ denotes the angular momentum of the $i$-th black hole. Unlike the
quasi-local charge~\eqref{eq:charge}, it is not straightforward to define the
quasi-local angular momentum of each black holes in general; see the review
article \cite{Szabados04} on the problem of quasi-local mass and angular
momentum. However, in the case of axial symmetry, there is a natural
definition: the Komar integral. In that case, we can ask whether inequality
\eqref{eq:3} holds. Finally, we can also combine, using the
Kerr-Newman black hole solution, both inequalities to obtain the general
inequality with charge and angular momentum. Namely, define the $m_i$ to be
\begin{equation}
  \label{eq:6}
  m_i= \sqrt{\frac{1}{4}\left(R_i +\frac{Q_i^2}{R_i}\right)^2
    +\frac{J^2_i}{R_i^2}}. 
\end{equation}

While it is still an open problem, inequality~\eqref{eq:3} for one
black hole in axial symmetry
is expected to be hold; see
the discussion in \cite{Mars:2009cj} and \cite{dain10d}.
However, if we assume that there are two or more black holes, our counterexample is relevant 
since when we set the charges and the angular momentum to zero, all these inequalities imply:
\begin{equation}   
\label{eq:GibbonsNoCharge}
    m \geq \frac12 \sum_i R_i,
\end{equation}
which, our example violates.  This inequaity  is stronger than the usual Riemannian Penrose inequality \cite{Bray}
\[
    m\geq \frac12 \left(\sum_i R_i^2\right)^{1/2}.
\]

As mentioned above, and as pointed out also in \cite{Gibbons72, Weinstein05},
the natural candidate to violate inequality \eqref{eq:GibbonsNoCharge} is a
configuration of two Schwarzschild black holes separated by a large
distance. This is precisely the counterexample we present.  Although all the
ingredients used in our argument have been present in the literature, it has
not been pointed out before to the best of our knowledge.  We believe that this
counterexample is important because it sheds some light on the quasi-local
aspect of Riemannian Penrose inequalities.

\section{Counterexample}
The Riemannian manifold that violates the inequality \eqref{eq:GibbonsNoCharge}  proposed by Gibbons \cite{Gibbons84} is
the well known Brill-Lindquist data \cite{Brill63}, which is a conformally flat
time-symmetric vacuum data defined on the differentiable manifold $M:=\R^3
\backslash \{x_1, x_2\}$ with the metric $h_{ij}=\phi^4 \delta_{ij}$ with the
conformal factor
\[
\phi = \left( 1 + \frac{\mu_1}{2|x - x_1|} + \frac{\mu_2}{2|x - x_2|} \right).
\] 
``Time-symmetric" here means that the second fundamental form of the three manifold
in the spacetime that is a solution to the Einstein equation vanishes. As
$\phi$ is harmonic on $\R^3$, the scalar curvature of the metric is zero. For
the sake of simplicity, we make the assumption $\mu_1=\mu_2=:\mu$, and 
$x_1 = (0,0,1)$, $x_2=(0,0,-1)$. 

The manifold $(M, h)$ has three asymptotically flat ends, namely $E_0$ where
$|x-x_1|, |x-x_2| \rightarrow \infty$, $E_1$ where $x \rightarrow x_1$ and
$E_2$ on which $x \rightarrow x_2$.  The end $E_0$ has mass $m(E_0)=\mu_1 +
\mu_2 = 2\mu$.  Let $\Omega_r=\{|x-x_1|>1/r, |x-x_2|>1/r, |x|<r\}$ and let $S_{r,i}=\{|x-x_i|=1/r\}$ for $i=1,2$, and
$S_0=\{|x|=r\}$, then for $r$ large enough, the boundary $S_{r,0}\cup S_{r,1}\cup S_{r,2}$ of $\Omega_r$ has positive
mean curvature with respect to the outer normal. Hence, by minimizing area over all surfaces enclosing $S_1$ and $S_2$
and enclosed in $S_0$, one can show that there is a surface of least area $\Sigma$ enclosing $E_1$ and
$E_2$, see~\cite[Theorem 1', p.~645]{meeks-simon-yau}. Suppose that
$\Sigma=\Sigma_1\cup\Sigma_2$ where $\Sigma_i$ is a compact minimal surface which encloses $E_i$, and
$A(\Sigma_i)=A_i$. In this situation, following Gibbons \cite{Gibbons72}, we have a lower bound $A_i > 16 \pi \mu^2$,
or equivalently in terms of the area radius $R_i>2\mu$.
Indeed, letting
\[
  \psi = \Big( 1 + \frac{\mu}{2|x - x_1|}  \Big),
\]
then $\bar h=\psi^4\delta_{ij}$ is a Schwarzschild metric, 
$\bar \Sigma = \{ |x-x_1| = \mu/2 \}$ minimizes area in $\bar h$ among all surfaces enclosing $E_1$, $\phi>\psi$, and
hence:
\[
  4\pi R_1^2=A_1 = A_h(\Sigma_1) > A_{\bar h} (\Sigma_1) \geq A(\bar\Sigma) = 16\pi\mu^2.
\]
Similarly, $R_2>2\mu$, and it follows immediately that $\frac12 (R_1+R_2) > 2\mu = m(E_0)$
violating\eqref{eq:GibbonsNoCharge}.

The next proposition shows that
for $\mu >0$ sufficiently small, there is no connected minimal sphere enclosing both
$E_1$ and $E_2$, showing that for small values of
$\mu>0$, the Brill-Lindquist data indeed provides a counterexample to
\eqref{eq:GibbonsNoCharge}.

\begin{prop}
  For sufficiently small $\mu > 0$, the Brill-Lindquist initial data described
  above contain no closed connected minimal surface enclosing both punctures   $x_1$ and $x_2$. 
\end{prop}

We remark that the statement follows from a direct application of Theorem 3.2
in \cite{Chrusciel-Mazzeo}.  For the sake of completeness, we present the proof
below. Our situation at hand is simpler than that of
\cite{Chrusciel-Mazzeo}, and the proof below illustrates the geometry of the
Brill-Lindquist metric concisely; see also~\cite{CCI} for a similar argument.

We also remark that this conclusion has been previously claimed based on
numerical evidence presented in \cite{Brill63} and \cite{Gibbons72}, and since then
 extensively numerically confirmed in the literature. Note that $\mu \to
0$ is equivalent to $L\to \infty$ while $\mu$ is kept constant, where $L$ is
the separation distance between the punctures (with respect to the flat
background metric).  This limit can be interpreted as the Newtonian limit of
the initial data. The physical content of this lemma is that in this limit the
initial data correspond to two separated black holes.

\begin{proof}
  To show the claim indeed holds, suppose $\Sigma$ is such a surface.  Without
  loss of generality, we can assume that $\Sigma$ is the outermost minimal
  surface for the end $E_0$.

  The configuration forces $\Sigma$ to intersect nontrivially with the $xy$-plane
  $\{z =0 \}$. Let $p$ be a point in $\Sigma \cap \{ z = 0\}$ and define the
  surface $\Sigma_1=\Sigma \cap B_{r} (p)$, with some fixed $r<1$. Here $B_{r}
  (p)$ denotes the (Euclidean) ball of radius $r$ centered at the point $p$.  By construction, $\Sigma_1$ is a minimal
surface, disjoint from the punctures, with
  nonempty boundary as the surface $\Sigma$ has to enclose both $x_1$ and
  $x_2$.  Since, by assumption, $\Sigma$ is outermost, it is a stable minimal
  surface. 

The following two consequences are then clearly in contradiction, implying that there is no such surface $\Sigma$.

\begin{enumerate}
\renewcommand{\labelenumi}{\roman{enumi})}
\item
By the Penrose inequality, the area of the surface $\Sigma$ (and hence the area
of $\Sigma_1$) is bounded above by the total mass of the data (which in our
configuration is given by $2\mu$);
\begin{equation}
  \label{eq:5}
  2\mu \geq
  \sqrt{\frac{A(\Sigma)}{16 \pi}} \geq  \sqrt{\frac{A(\Sigma_1)}{16 \pi}}
\end{equation}
And hence, we have that  $A(\Sigma_1)\to 0$ as $\mu \to 0$. 
\item
We have a lower bound on the area $A(\Sigma_1)$ independent of $\mu$.
This follows from an estimate of the sup norm of the second fundamental form
as in Theorem 2 of~\cite{Schoen83c} for stable minimal surfaces.  
It gives  that the norm of the second fundamental form of $\Sigma_1$ is bounded uniformly (in $\mu$) by some positive
constant $C$ (which depends on $0<r<1$ ), provided $\mu$ is
sufficiently small.   Then, there exists a sufficiently small $\varepsilon > 0$ independent of $\mu$, so that the
surface $\Sigma \cap B_{\varepsilon} (p)$ can be described as a graph over the tangent plane $T_p \Sigma$, which in turn
gives a positive lower bound of the
area of $\Sigma \cap B_{\varepsilon} (p)$ independent of $\mu$.
\end{enumerate}

\end{proof}

\section*{Acknowledgments}

This work began at the conference ``GR19, International Society on General
Relativity \& Gravitation 19th International Conference'', Mexico,
July 5-9, 2010.  S. D. and S. Y.  would like to thank the organizers of this
conference for the invitation.

S. D. is supported by CONICET (Argentina).  This work was supported in part by
grant PIP 6354/05 of CONICET (Argentina), grant 05/B415 Secyt-UNC (Argentina)
and the Partner Group grant of the Max Planck Institute for Gravitational
Physics, Albert-Einstein-Institute (Germany).
 
S.Y is partially supported by Grant-Aid for Scientific Research 
(No.~20540201).


\end{document}